\documentclass[12pt]{article}

\usepackage{amsmath,amssymb,graphicx,color,array,cite}
\numberwithin{equation}{section}

\evensidemargin \oddsidemargin

\begin{document}

\newcommand{\story}{\vspace{5mm} \noindent $\spadesuit$ }

\begin{titlepage}

\renewcommand{\thefootnote}{\fnsymbol{footnote}}


\begin{flushright}
CQUeST-2008-0174 \\
\end{flushright}

\vspace{15mm}
\baselineskip 9mm
\begin{center}
  {\Large \bf Anomaly and Hawking radiation \\ 
  from regular black holes}
\end{center}

\baselineskip 6mm
\vspace{10mm}
\begin{center}
  Wontae Kim\footnote{\tt wtkim@sogang.ac.kr}
  \\[3mm]
  {\sl Department of Physics and Center for Quantum Spacetime \\
    Sogang University, C.P.O. Box 1142, Seoul 100-611, South Korea}
  \\[10mm]
  Hyeonjoon Shin\footnote{\tt hshin@sogang.ac.kr} and
  Myungseok Yoon\footnote{\tt younms@sogang.ac.kr}
  \\[3mm] 
  {\sl Center for Quantum Spacetime, Sogang University, Seoul
    121-742, South Korea }
  \\[3mm]

\end{center}

\thispagestyle{empty}

\vfill
\begin{center}
{\bf Abstract}
\end{center}
\noindent
We consider the Hawking radiation from two regular black holes, the 
minimal model and the noncommutative black hole.  
The flux of Hawking radiation is derived by applying the anomaly 
cancellation method proposed by Robinson and Wilczek. Two regular
black holes have the same radiation pattern except for the detailed
expression for the Hawking temperature. The resulting flux of the 
energy-momentum tensor is shown to be precisely the same with the 
thermal flux from each regular black hole at the Hawking temperature.
\\ [5mm]
Keywords : Hawking radiation, anomaly, regular black hole
\\
PACS numbers : 04.62.+v, 04.70.Dy, 11.30.-j

\vspace{5mm}
\end{titlepage}

\baselineskip 6.6mm
\renewcommand{\thefootnote}{\arabic{footnote}}
\setcounter{footnote}{0}

\section{Introduction}

Hawking radiation  is the quantum effect of fields in a classical 
space-time background with an event horizon\cite{hawking}, and 
provides an key ingredient to understand the nature of black hole 
horizon.  Since the quantum effect of gravity itself becomes no 
longer negligible near the black hole horizon, Hawking radiation 
also provides a basic information in formulating the theory of 
quantum gravity.  Because of its importance, it may be useful to
have various interpretations from various different angles, which
may lead to some breakthrough in understanding the nature of black 
hole.  

Recently, Robinson and Wilczek \cite{rw} came up with a new 
interpretation about the Hawking radiation.  Their proposal is
that the Hawking radiation plays the role of preserving general 
covariance at the quantum level by canceling the diffeomorphism 
anomaly at the event horizon.  It should be noted that the
proposal is supposed to be valid in any space-time dimension, 
contrary to the previous similar work \cite{cf} which is restricted 
to two-dimensional space-time.

For their formulation, Robinson and Wilczek considered the static and
spherically symmetric black hole.  Elaboration of the original idea 
and the extensions to more general black holes, the charged and 
rotating black holes, have been done in \cite{iso1,iso2},
where it has been shown that Hawking radiation is capable of
canceling anomalies of local symmetries at the horizon.  
In subsequent works \cite{murata}-\cite{wu}, the method of anomaly 
cancellation has been 
applied to various black objects in various dimensions even including 
the black ring.  (For a review, see Ref.\ \cite{das}.)
As some further elaborations, Hawking fluxes of 
higher-spin currents have been studied in \cite{iso4} and some
clarifications on the use of anomaly have been given in \cite{baner}.
All the results until now have given the expected
Hawking fluxes and so put the validity of the method on a firmer
footing.  
 
In this paper, we consider another interesting class of black
hole, the regular black hole, and study the Hawking radiation from
it through the anomaly cancellation method.  Let us note that one kind 
of regular black hole has been already explored \cite{Setare}.  It is 
the well-known BTZ black hole in three space-time dimensions.  What
we are concerned about is the four-dimensional regular black hole.
In fact, we consider two regular black holes, which are the minimal
model \cite{hayward} and the noncommutative black hole \cite{nss}.  
The noncommutative
black hole is especially attractive, since it has a noncommutative 
parameter and how its effect appears in the Hawking radiation is
an interesting issue.  Although two regular black holes are surely 
different to each other, as we will see, these can be dealt with 
almost simultaneously.

The organization of this paper is as follows: In the next section, 
two regular black holes are described.  We consider a test 
real scalar field in the regular black hole background in Sec.~\ref{qf}, 
and show that, near the horizon, the action for the scalar field 
reduces to a two-dimensional theory in a certain background.  
In Sec.~\ref{fluxes}, the flux of the energy-momentum 
tensor is derived by applying the method of anomaly cancellation 
to the effective two-dimensional theory, and is shown to be the
same with the thermal flux at the Hawking temperature.  Based on the
result, the difference between two regular black holes is discussed.

\section{Regular black holes}
\label{sec:RBH}

In this section, we introduce two regular black holes, the minimal
model and the noncommutative black hole, and explain their basic 
peculiar properties.

Two regular black holes considered here have the common structure
that the geometry is static and spherically symmetric,
\begin{equation}
  \label{metric:RBH}
  ds^2 = - f(r) dt^2 + f(r)^{-1} dr^2 + r^2 d\Omega_2^2 ~,
\end{equation}
where $f(r)$ is a function which vanishes at the event horizon.
The explicit form of the function $f(r)$ distinguishes the minimal 
model and the noncommutative black hole, whose descriptions are
given below in order.

\subsection{Minimal model black hole}
\label{ssec:minimal}

Let us suppose a metric which behaves like Schwarzschild black hole
at infinity as follows
\begin{equation}
  \label{f:inf}
  f(r) \sim 1 - \frac{2m}{r} \quad \mathrm{as} ~ r \to \infty,
\end{equation}
where $m$ is the total mass.  As for the behavior near the origin or
the center, we require the flatness.  This may lead to the following 
choice
\begin{equation}
  \label{f:center}
  f(r) \sim 1 - \frac{r^2}{\ell^2} \quad \mathrm{as} ~ r \to 0,
\end{equation}
where $\ell$ is a positive constant.  Actually, this is not an
ad hoc choice but is the solution of the Einstein equation with
the cosmological constant, $G_{\mu\nu} = -\Lambda g_{\mu\nu}$, 
near the origin, when the cosmological constant $\Lambda$ is chosen 
as $\Lambda = 3/\ell^2$.

\begin{figure}[pt]
  \includegraphics[width=0.5\textwidth]{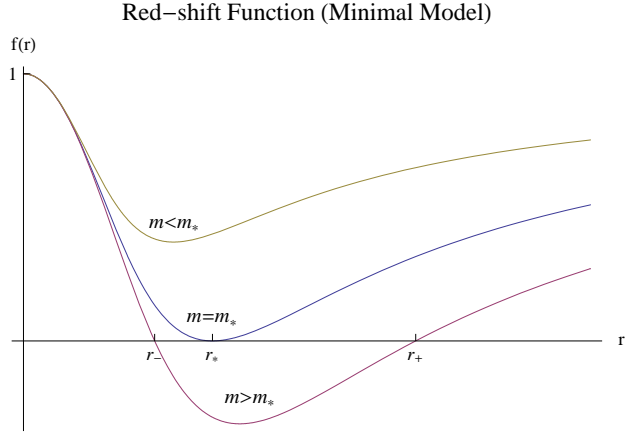}
  \caption{The red-shift function $f(r)$ is regular for all $r$.
    For $m>m_*$ (bottom), there exists two horizons. When
    $m=m_* = 3\sqrt{3}\ell/4$ (middle), they degenerate.
    In the case where $m<m_*$ (top), there is no horizon.} 
  \label{fig:metric:mbh}
\end{figure}

One possible interpolating function, which asymptotes to (\ref{f:inf}) 
and (\ref{f:center}) in the corresponding regions, has been derived in 
\cite{hayward}, and its explicit expression is
\begin{equation}
  \label{metric:mbh}
  f(r) = 1 - \frac{2mr^2}{r^3 + 2\ell^2 m} ~.
\end{equation}
The geometry (\ref{metric:RBH}) with the function $f$ of 
(\ref{metric:mbh}) has two horizons, the inner horizon $r_-$ and 
the outer horizon $r_+$, which satisfy the inequality 
$r_- \le r_* \le r_+$.  Here $r_*$ is a constant defined
by $r_* \equiv \sqrt{3} \ell$.  Because of the presence of the
event horizons, we now have a black hole, called the minimal model
black hole.  The total mass $m$ of the black hole is expressed
in terms of the horizons as  
\begin{equation}
  \label{m:mbh}
  m = \frac{r_\pm^3}{2 (r_\pm^2 - \ell^2)} ~,
\end{equation}
and one can see that it has the minimum value $m_*=3\sqrt{3}\ell/4$ 
when $r_+ = r_*$.  

\begin{figure}[pt]
  \includegraphics[width=0.5\textwidth]{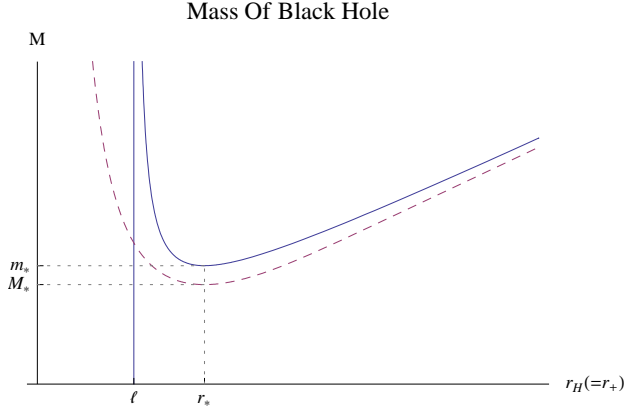}
  \caption{Black hole mass versus the position of the event horizon.
        The solid and the dashed lines correspond to
        the minimal model and the noncommutative black holes, 
        respectively. There is minimal mass $m_*$ ($M_*$) to form 
        minimal model black hole (noncommutative black hole).
        Each black hole has the minimal event horizon for the minimal
        mass. The minimal event horizons are given by 
        $r_*=\sqrt{3}\ell$ for the minimal model black hole and by
        $r_*=2\alpha\sqrt{\theta}$ for the
        noncommutative black hole, respectively. In this figure, 
        the minimal horizons are adjusted to have the same value
        for comparison.
      }
  \label{fig:M}
\end{figure}

We note that the minimal model black hole exists
only for $m \ge m_*$ as illustrated in Fig.~\ref{fig:metric:mbh}.
If $m$ is less than $m_*$, we do not see any horizon and thus the 
metric (\ref{metric:RBH}) does not represent a black hole.  One
important point is that $r_*$ is the minimal value of the event
horizon of the minimal model regular black hole.  Therefore,
one may call it the minimal event horizon.  The presence of the
minimal horizon is also the case for the noncommutative black hole.
In Fig.~\ref{fig:M}, we illustrate the existence of the minimal
mass by plotting the black hole mass with respect to
the event horizon.  This gives another way of realizing that
a black hole does not form if $m < m_*$.

\subsection{Noncommutative black hole}
\label{ssec:NBH}

We now turn our attention to the noncommutative black hole.
It has been shown that noncommutativity eliminates point-like
structures in favor of smeared objects in flat
spacetime~\cite{ss}. The effect of smearing is mathematically
implemented by replacing the Dirac-delta function in position space 
with a Gaussian distribution of the width $\sqrt{\theta}$. 
In a static and spherically symmetric case,
the mass density of a gravitational source is chosen to be \cite{nss}
\begin{equation}
  \label{rho}
  \rho_\theta = \frac{M}{(4\pi\theta)^{3/2}} \exp \left(
    -\frac{r^2}{4\theta} \right) ~,
\end{equation}
where $\theta$ is a constant parameter representing
noncommutativity and $M$ is the total mass.  This mass density
implies that the total mass is diffused over the region 
of linear size $\sqrt{\theta}$. 

For a static and spherically symmetric geometry, the 
energy-momentum tensor is given by 
${T^\mu}_\nu = {\rm diag} (-\rho_\theta, p_r, p_\bot, p_\bot)$.
From the conservation law, the radial and the tangential pressure 
are related to the mass density as $p_r = -\rho_\theta$ and 
$p_\bot = -\rho_\theta - \frac12 r \partial_r \rho_\theta$, respectively. 
Then, from the Einstein equation, we obtain the line element
(\ref{metric:RBH}) 
with the function 
\begin{equation}
  \label{f}
  f(r) = 1 - \frac{4M}{r\sqrt{\pi}} \gamma\left( \frac32,
    \frac{r^2}{4\theta} \right) ~,
\end{equation}
where $\gamma$ is the lower incomplete gamma function,
\begin{equation}
  \label{gamma}
  \gamma\left(a, z \right) \equiv \int_0^z t^{a-1} e^{-t} dt ~.
\end{equation}
The function of (\ref{f}) leads to the metric for the noncommutative 
black hole.

\begin{figure}[pt]
  \includegraphics[width=0.5\textwidth]{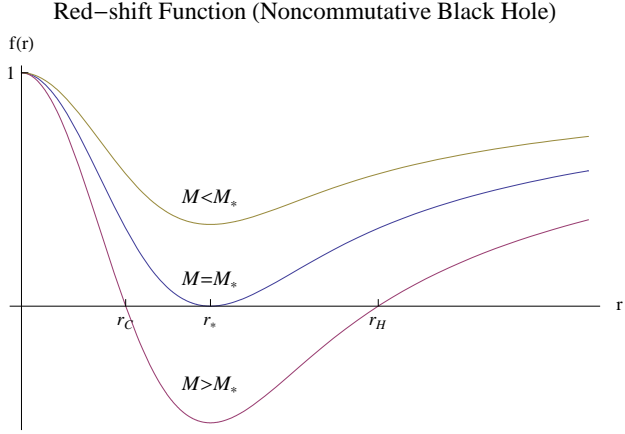}
  \caption{The red-shift function $f(r)$ is regular for all $r$ and
    its minimum value always appears at $r=r_*$. Like the case of the
    minimal model black hole, there is no horizon
    for $M<M_*$ (top), while two horizons exist for 
    $M>M_*$ (bottom) and degenerated one for
    $M=M_*\approx1.9\sqrt{\theta}$ (middle), respectively.} 
  \label{fig:metric:NBH}
\end{figure}

Like the case of the minimal model black hole, we see that there are
two horizons, that is, the inner (Cauchy) horizon $r_C$ and the
outer (event) horizon $r_H$, and there exists the minimal mass $M_*$
below which no black hole can be formed. These are illustrated in
Fig.~\ref{fig:metric:NBH}.  

Let us now evaluate the minimal mass.
First of all, from $f(r_H)=0$, we obtain the relation between the
total mass and the event horizon as follows:
\begin{equation}
  \label{M}
  M = \frac{r_H \sqrt{\pi}}{4\gamma_H} ~,
\end{equation}
where $\gamma_H = \gamma\left(\frac32, \frac{r_H^2}{4\theta} \right)$.
Based on this relation, we can plot a graph shown in Fig.~\ref{fig:M},
and see the presence of the minimal mass $M_*$.  We hope to
have the explicit expression of $M_*$.  However, the analytical
evaluation is not an easy task because the relation (\ref{M}) contains
an incomplete gamma function.  So we take the numerical evaluation of 
$M_*$.  Before doing that, it is convenient to hide the noncommutativity
parameter $\theta$.  This is accomplished by 
redefining quantities in the red-shift function $f(r)$ as 
$M \to M'=M/(2\sqrt\theta)$ and $r \to r'=r/(2\sqrt\theta)$ such that
$f(r)$ becomes $1 - 4M'\gamma(3/2,r'^2)/(r'\sqrt\pi)$. 
Then, the redefined red-shift function leads us to have
$M'=r'_H\sqrt\pi/[4\gamma(3/2,{r'}_H^2)]$.  By solving this and 
returning back to the original quantities, we get 
$r_* = 2 \alpha \sqrt{\theta}$ and
$M_* = \sqrt{\pi\theta}/(4\alpha^2e^{-\alpha^2})$, 
where $\alpha \equiv r'_*$ is purely a numerical constant 
determined by
\begin{equation}
  \label{alpha}
  2\alpha^3 e^{-\alpha^2} = \gamma\left(\frac32, \alpha^2\right) ~.
\end{equation}
A numerical evaluation gives $\alpha \approx 1.51122$, and thus we 
finally get
$r_* \approx3.02244\sqrt\theta$ and $M_* \approx1.90412\sqrt\theta$.

\section{Quantum field near the horizon}
\label{qf}

We now consider a real free scalar field in the regular black
hole background, Eq.~(\ref{metric:RBH}), and investigate its action near 
the horizon.  Although we are dealing with two regular black holes 
characterized by functions (\ref{metric:mbh}) and (\ref{f}), the
difference between them is not important in this section.  It is sufficient
to know that both of the regular black holes have the same metric form.

The action for the real scalar field $\varphi$ in the background,
Eq.~(\ref{metric:RBH}), is evaluated as
\begin{align}
  S[\varphi] &= - \int d^4 x \sqrt{-g} g^{\mu\nu} \partial_\mu
     \varphi \partial_\nu \varphi \notag \\
    &= \int dt dr \ r^2 \int d\vartheta d\phi\, \sin\vartheta \ \varphi
    \left[ -\frac{1}{f} \partial_t^2 +
      \frac{1}{r^2} \partial_r (r^2f\partial_r ) + \frac{1}{r^2}
      \nabla_\Omega^2 \right] \varphi,  
\label{S:4d}
\end{align}
where $\nabla^2_\Omega$ denotes the Laplacian on unit two sphere.
If we perform a wave decomposition of $\varphi$ in terms
of spherical harmonics
$\varphi = \sum_\ell \varphi_\ell Y_\ell(\vartheta,\phi)$, 
where $\ell$ is the collection of angular quantum numbers
of the spherical harmonics and $\varphi_\ell$ depends on the coordinates,
$t$ and $r$, then we see that the action is reduced to a
two-dimensional effective theory with an infinite collection of fields
labeled by $\ell$.  Next, in order to see what happens near the horizon,
it is helpful to take a transformation to the tortoise coordinate
$r^*$, which, in our case, is defined by
$\partial r^* / \partial r = 1 / f(r)$, 
and leads to $\int dr = \int dr^* f(r(r^*))$.  If we now go to the region
near the horizon, the factor $f(r(r^*))$ appears to be a 
suppression factor vanishing exponentially fast, and thus the terms 
in the action which do not have some appropriate factor compensating it 
can be ignored.  In the present case, one can easily see that
the terms coming from the Laplacian on unit two sphere are
suppressed by $f(r(r^*))$.  We note that the suppression also takes
place for the mass term or the interaction terms of $\varphi$ when
they are included in the action (\ref{S:4d}).  

After all, the action near the horizon becomes
\begin{equation}
  \label{S:2d}
  S[\varphi] = \sum_\ell \int dt dr \ r^2 \varphi_\ell \left[
    -\frac{1}{f} \partial_t^2 + \frac{1}{r^2} \partial_r
    (r^2f\partial_r ) \right] \varphi_\ell ~. 
\end{equation}
One can check that this action describes an infinite set of 
massless two-dimensional scalar fields in the following background:
\begin{eqnarray}
  ds^2 &=& -f(r) dt^2 + \frac{1}{f(r)} dr^2 ~, \nonumber \\
  \Phi &=& r^2 ~, 
\label{2dbg}
\end{eqnarray}
where $\Phi$ is the two-dimensional dilaton field.

\section{Anomalies and Hawking fluxes}
\label{fluxes}

In this section, having the two-dimensional effective theory
near the horizon (\ref{S:2d}) and the background (\ref{2dbg}), we are 
going to consider the problem of Hawking radiation following
the approach based on the anomaly cancellation proposed in
\cite{rw,iso1}.  Before starting, we would like to note that the two 
regular black holes distinguished by different $f$'s,
(\ref{metric:mbh}) and (\ref{f}), can be treated simultaneously.
However, the final results will, of course, turn out to be different.

The anomaly approach of \cite{rw} begins with an observation 
that, since the horizon is a null hypersurface, all ingoing
(left moving) modes at the horizon can not classically affect physics
outside the horizon.  This implies that they may be taken to be out of
concern at the classical level and thus the effective two-dimensional
theory becomes chiral, that is, the theory only of outgoing (right
moving) modes.  If we now perform the path integration of right moving
modes, the resulting quantum effective action becomes anomalous under
the general coordinate transformation, due to the absence
of the left moving modes. However, such anomalous behaviors are in
contradiction to the fact that the underlying theory is not anomalous.
The reason for this is simply that we have ignored the quantum effects
of the classically irrelevant left moving modes at the horizon.  Thus
anomalies must be cancelled by including them. 

The above argument implies that anomaly is localized at the horizon
$r_H$.  In order to avoid some possible difficulties due to the sharp
localization of the anomaly, it is convenient to regard the quantum 
effective action to be anomalous in an infinitesimal slab, 
$r_H \le r \le r_H + \epsilon$, which is the region near the horizon.  
The limit $\epsilon \rightarrow 0$ is taken at the end of the 
calculation.  This leads to
a splitting of the region outside the horizon, $r_H \le r \le \infty$,
into two regions, $r_H \le r \le r_H + \epsilon$ and $r_H + \epsilon
\le r \le \infty$.  Then there will be the gravitational anomaly
near the horizon, $r_H \le r \le r_H + \epsilon$.

What we are interested in is the problem of determining the flux of the
energy-momentum tensor through the cancellation of the gravitational
anomaly.  Since the region outside the horizon has been divided into
two regions, we first write the energy-momentum tensor as a sum
\begin{align}
T^\mu_\nu = T^\mu_{\nu(o)}\Theta_+(r) +  T^\mu_{\nu(H)} H(r) ~,
\label{tsplit}         
\end{align}
where $\Theta_+(r) = \Theta(r-r_H-\epsilon)$ and $H(r)=1-\Theta_+(r)$.
Among the components of the energy-momentum tensor, only the flux
in the radial direction, $T^r_t$, is of concern to us.
Apart from the near horizon region, $r_H + \epsilon \le r \le \infty$, 
it is conserved
\begin{align}
\partial_r T^r_{t(o)} = 0 ~.
\label{toeq}
\end{align}
On the other hand, in the near horizon region,
$r_H \le r \le r_H + \epsilon$, we have
anomalous conservation equation \cite{rw} as
\begin{align}
\partial_r T^r_{t(H)} =  \partial_r N^r_t ~,
\label{theq}
\end{align}
where $ N^r_t =( f^{\prime 2}+f f^{\prime\prime})/192\pi$.  (The prime
denotes the derivative with respect to $r$.)  
The non-vanishing term in the right-hand side is due to the 
gravitational anomaly for the consistent
energy-momentum tensor \cite{aw}.  Now it is not a difficult task to
integrate Eqs.~(\ref{toeq}) and (\ref{theq}) and obtain
\begin{align}
T^r_{t(o)} &= a_o  ~, \notag \\
T^r_{t{(H)}} &= a_H + \int^r_{r_H} dr \partial_r  N^r_t ~,
\label{tsol}
\end{align}
where $a_o$ and $a_H$ are integration constants. Here $a_o$ is the
energy flux which is of our concern.

Next, we consider the variation of quantum effective action $W$ under
a general coordinate transformation in the time direction with a
transformation parameter $\xi^t$:
\begin{align}
- \delta W 
&= \int d^2x \sqrt{-g} \; \xi^t \nabla_\mu T^\mu_{t} 
\notag \\
&= \int d^2x \; \xi^t
  \left[ 
        \partial_r \left( N^r_t  H \right)       
    + \left( T^r_{t~(o)} - T^r_{t~(H)} + N^r_t
      \right) \delta(r-r_H -\epsilon) 
     \right] ~,
\label{gvar}
\end{align} 
where Eqs.~(\ref{tsplit}), (\ref{toeq}) and (\ref{theq}) have been
used for obtaining the second line.
As mentioned before, the full quantum effective action of the 
underlying theory must be diffeomorphism invariant.  
The full effective action includes the quantum effects of the ingoing 
modes near the horizon, whose variation under the general coordinate
transformation gives a term canceling the first term of (\ref{gvar}).  
For the general covariance of the full quantum effective action, the 
coefficient of the delta function in
Eq.~(\ref{gvar}) is also required to vanish. This
requirement leads us to have the following relation.
\begin{align}
a_o = a_H - N^r_t(r_H) ~,
\end{align}
where the solution Eq.~(\ref{tsol}) has been used.  For determining
$a_o$, the value of the energy-momentum flux at the horizon, $a_H$, 
should be fixed.  This is done by
imposing a condition that the covariant energy-momentum tensor
vanishes at the horizon for regularity at the future horizon
\cite{iso2}.  Then, from the expression of the covariant
energy-momentum tensor \cite{bz,bk}, $\tilde{T}^r_t = T^r_t
+\frac{1}{192\pi} (f f'' -2(f')^2)$, the condition
$\tilde{T}^r_t(r_H)=0$ gives
\begin{align}
a_H= \frac{\kappa^2}{24 \pi} = 2N^r_t(r_H) ~,
\end{align}
where $\kappa$ is the surface gravity at the horizon,
\begin{align}
\kappa = \frac{1}{2} \partial_r f |_{r=r_H} ~.
\label{sg}
\end{align}
Having the expression of $a_H$, the
flux of the energy-momentum tensor is finally determined as
\begin{align}
a_o = N^r_t(r_+) = \frac{\pi}{12} T_H^2 ~,
\label{tflux}
\end{align}
where the relation between the Hawking temperature 
$T_H$ and the surface gravity, $T_H = \kappa / 2\pi$, has been used.
This is precisely the thermal flux from a regular black hole at the
Hawking temperature.

\begin{figure}[pt]
  \includegraphics[width=0.5\textwidth]{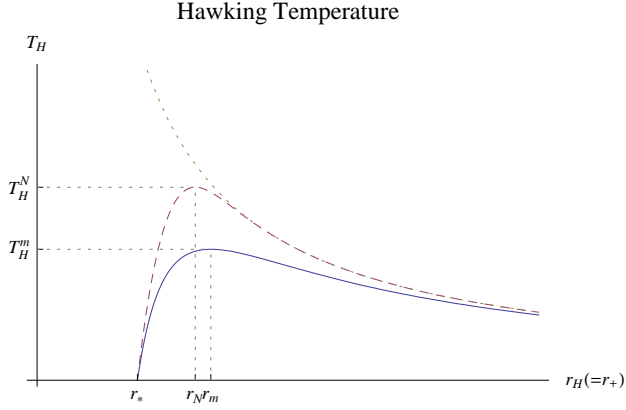}
  \caption{Behavior of Hawking temperature in terms of the position of
    event horizon.  The solid, dashed, and dotted lines
    correspond to the minimal model, the noncommutative black hole, 
    and the Schwarzschild black hole, respectively. 
    The minimal model (noncommutative) black hole has the maximum 
    Hawking temperature $T_H^m$ ($T_H^N$) at $r_+=r_m$ ($r_H = r_N$). 
    One can check that it is the same with that of the
    Reissner-Nordstr\"om black hole with the charge $Q=\sqrt{3}\ell$.} 
  \label{fig:T}
\end{figure}
Because the two regular black holes have the same metric form, the
expression for the flux of the energy-momentum tensor (\ref{tflux})
is valid for both of them.  However, as mentioned at the 
beginning in this section, the final results are different for the 
two black holes.  Let us evaluate the Hawking temperature for each
black hole based on (\ref{sg}).  As for the minimal black hole, 
by using the function $f$ in Eq.~(\ref{metric:mbh}), it is
obtained as
\begin{equation}
\label{kappa:mbh}
  T_H = \frac{r_+^2 - 3\ell^2}{4\pi r_+^3} ~.
\end{equation}
Intriguing thing is that this is the same as that of the 
Reissner-Nordstr\"om black hole when the electric charge of the 
black hole is given 
by $Q=\sqrt{3}\ell$.  If we turn to another black hole, the 
noncommutative black hole, the function $f$ in Eq.~(\ref{f}) leads 
us to
\begin{equation}
\label{TH}
  T_H  =  \frac{1}{4\pi\, r_H\! } \left[1 - \frac{M
      r_H^2}{\sqrt{\pi}\theta^{3/2}} \exp\left(
      -\frac{r_H^2}{4\theta}\right) \right] ~.
\end{equation}
We see that two Hawking temperatures are
obviously different and thus lead to the different expressions for 
the flux.  However, we would like to note that they have the
similar pattern in terms of the position of the event horizon as 
shown in Fig.~\ref{fig:T}.  Such pattern may be argued to be
a characteristic feature of regular black holes, which is 
distinguished from, for example, that of the Schwarzschild black 
hole.

\section*{Acknowledgments}
We would like to thank B-H.\ Lee for useful discussions.
This work was supported by the Science Research Center Program of the
Korea Science and Engineering Foundation through the Center for
Quantum Spacetime (CQUeST) of Sogang University with grant number
R11-2005-021.


\begin{thebibliography}{10}
\bibitem{hawking}
  S.~W.~Hawking,
  ``Particle Creation By Black Holes,''
  Commun.\ Math.\ Phys.\  {\bf 43} (1975) 199
  [Erratum-ibid.\  {\bf 46} (1976) 206].
\bibitem{rw}
  S.~P.~Robinson and F.~Wilczek,
  ``A relationship between Hawking radiation and gravitational 
  anomalies,''
  Phys.\ Rev.\ Lett.\  {\bf 95} (2005) 011303
  [arXiv:gr-qc/0502074].
\bibitem{cf}
  S.~M.~Christensen and S.~A.~Fulling,
  ``Trace Anomalies And The Hawking Effect,''
  Phys.\ Rev.\  D {\bf 15} (1977) 2088.
\bibitem{iso1}
  S.~Iso, H.~Umetsu and F.~Wilczek,
  ``Hawking radiation from charged black holes via gauge and gravitational
  anomalies,''
  Phys.\ Rev.\ Lett.\  {\bf 96} (2006) 151302
  [arXiv:hep-th/0602146].
\bibitem{iso2}
  S.~Iso, H.~Umetsu and F.~Wilczek,
  ``Anomalies, Hawking radiations and regularity in rotating black holes,''
  Phys.\ Rev.\  D {\bf 74} (2006) 044017
  [arXiv:hep-th/0606018].
\bibitem{murata}
  K.~Murata and J.~Soda,
  ``Hawking radiation from rotating black holes and gravitational  anomalies,''
  Phys.\ Rev.\  D {\bf 74} (2006) 044018
  [arXiv:hep-th/0606069].
\bibitem{Vagenas:2006qb}
  E.~C.~Vagenas and S.~Das,
  ``Gravitational anomalies, Hawking radiation, and spherically symmetric
  black holes,''
  JHEP {\bf 0610} (2006) 025
  [arXiv:hep-th/0606077].
\bibitem{Setare}
  M.~R.~Setare,
  ``Gauge and gravitational anomalies and Hawking radiation of rotating BTZ
  black holes,''
  Eur.\ Phys.\ J.\  C {\bf 49} (2007) 865
  [arXiv:hep-th/0608080];
  Q.~Q.~Jiang, S.~Q.~Wu and X.~Cai,
  ``Hawking radiation from (2+1)-dimensional BTZ black holes,''
  Phys.\ Lett.\  B {\bf 651} (2007) 58
  [arXiv:hep-th/0701048].
\bibitem{Xu:2006tq}
  Z.~Xu and B.~Chen,
  ``Hawking radiation from general Kerr-(anti)de Sitter black holes,''
  Phys.\ Rev.\  D {\bf 75} (2007) 024041
  [arXiv:hep-th/0612261].
\bibitem{Iso:2006xj}
  S.~Iso, T.~Morita and H.~Umetsu,
  ``Quantum anomalies at horizon and Hawking radiations in Myers-Perry black
  holes,''
  JHEP {\bf 0704} (2007) 068
  [arXiv:hep-th/0612286].
\bibitem{Jiang:2007gc}
  Q.~Q.~Jiang and S.~Q.~Wu,
  ``Hawking radiation from rotating black holes in anti-de Sitter spaces via
  gauge and gravitational anomalies,''
  Phys.\ Lett.\  B {\bf 647} (2007) 200
  [arXiv:hep-th/0701002].
\bibitem{Jiang:2007wj}
  Q.~Q.~Jiang, S.~Q.~Wu and X.~Cai,
  ``Hawking radiation from the dilatonic black holes via anomalies,''
  Phys.\ Rev.\  D {\bf 75} (2007) 064029
  [Erratum-ibid.\  {\bf 76} (2007) 029904]
  [arXiv:hep-th/0701235].
\bibitem{Kui:2007dy}
  X.~Kui, W.~Liu and H.~b.~Zhang,
  ``Anomalies of the Achucarro-Ortiz black hole,''
  Phys.\ Lett.\  B {\bf 647} (2007) 482
  [arXiv:hep-th/0702199].
\bibitem{Shin:2007gz}
  H.~Shin and W.~Kim,
  ``Hawking radiation from non-extremal D1-D5 black hole via anomalies,''
  JHEP {\bf 0706} (2007) 012
  [arXiv:0705.0265 [hep-th]].
\bibitem{Peng:2007pk}
  J.~J.~Peng and S.~Q.~Wu,
  ``Hawking radiation from the Schwarzschild black hole with a global monopole
  via gravitational anomaly,''
  arXiv:0705.1225 [hep-th];
  S.~Q.~Wu and J.~J.~Peng,
  ``Hawking radiation from the Reissner-Nordstr\'{o}m black hole with a
  global monopole via gravitational and gauge anomalies,''
  Class.\ Quant.\ Grav.\  {\bf 24} (2007) 5123
  [arXiv:0706.0983 [hep-th]];
  S.~Gangopadhyay,
  ``Hawking radiation in Reissner-Nordstr\'{o}m blackhole with a global
  monopole via Covariant anomalies and Effective action,''
  arXiv:0803.3492 [hep-th].
\bibitem{Jiang:2007mi}
  Q.~Q.~Jiang,
  ``Hawking radiation from black holes in de Sitter spaces,''
  Class.\ Quant.\ Grav.\  {\bf 24} (2007) 4391
  [arXiv:0705.2068 [hep-th]];
  Q.~Q.~Jiang, S.~Q.~Wu and X.~Cai,
  ``Anomalies and de Sitter radiation from the generic black holes in de
  Sitter spaces,''
  Phys.\ Lett.\  B {\bf 651} (2007) 65
  [arXiv:0705.3871 [hep-th]].
\bibitem{Chen:2007pp}
  B.~Chen and W.~He,
  ``Hawking Radiation of Black Rings from Anomalies,''
  arXiv:0705.2984 [gr-qc];
  U.~Miyamoto and K.~Murata,
  ``On Hawking radiation from black rings,''
  Phys.\ Rev.\  D {\bf 77} (2008) 024020
  [arXiv:0705.3150 [hep-th]].
\bibitem{kim}
  W.~Kim and H.~Shin,
  ``Anomaly Analysis of Hawking Radiation from Acoustic Black Hole,''
  JHEP {\bf 0707} (2007) 070
  [arXiv:0706.3563 [hep-th]].
\bibitem{Murata:2007zr}
  K.~Murata and U.~Miyamoto,
  ``Hawking radiation of a vector field and gravitational anomalies,''
  Phys.\ Rev.\  D {\bf 76} (2007) 084038
  [arXiv:0707.0168 [hep-th]].
\bibitem{Peng:2007kv}
  J.~J.~Peng and S.~Q.~Wu,
  ``Can the anomaly cancellation method derive a correct Hawking temperature of
  a Schwarzschild black hole in the isotropic coordinates ?,''
  arXiv:0709.0044 [hep-th].
\bibitem{Peng:2007nj}
  J.~J.~Peng and S.~Q.~Wu,
  ``Covariant anomaly and Hawking radiation from the modified black hole in the
  rainbow gravity theory,''
  arXiv:0709.0167 [hep-th].
\bibitem{Ma:2007xr}
  Z.~Z.~Ma,
  ``Hawking radiation of black p-branes via gauge and gravitational
  anomalies,''
  arXiv:0709.3684 [hep-th].
\bibitem{Gangopadhyay:2007fm}
  S.~Gangopadhyay and S.~Kulkarni,
  ``Hawking radiation in GHS and non-extremal D1-D5 blackhole via covariant
  anomalies,''
  Phys.\ Rev.\  D {\bf 77} (2008) 024038
  [arXiv:0710.0974 [hep-th]];
  S.~Gangopadhyay,
  ``Hawking radiation in GHS blackhole, Effective action and Covariant Boundary
  condition,''
  arXiv:0712.3095 [hep-th].
\bibitem{Huang:2007ed}
  C.~G.~Huang, J.~R.~Sun, X.~n.~Wu and H.~Q.~Zhang,
  ``Gravitational Anomaly and Hawking Radiation of Brane World Black Holes,''
  arXiv:0710.4766 [hep-th].
\bibitem{Peng:2008ru}
  J.~J.~Peng and S.~Q.~Wu,
  ``Covariant anomalies and Hawking radiation from charged rotating black
  strings in anti-de Sitter spacetimes,''
  arXiv:0801.0185 [hep-th].
\bibitem{Wu:2008yx}
  X.~n.~Wu, C.~G.~Huang and J.~R.~Sun,
  ``On Gravitational anomaly and Hawking radiation near weakly isolated
  horizon,''
  arXiv:0801.1347 [gr-qc].
\bibitem{wu}
  S.~Q.~Wu, J.~J.~Peng and Z.~Y.~Zhao,
  ``Anomalies, effective action and Hawking temperatures of a Schwarzschild
  black hole in the isotropic coordinates,''
  arXiv:0803.1338 [hep-th].
\bibitem{das}
  S.~Das, S.~P.~Robinson and E.~C.~Vagenas,
  ``Gravitational anomalies: a recipe for Hawking radiation,''
  arXiv:0705.2233 [hep-th].
\bibitem{iso4}
  S.~Iso, T.~Morita and H.~Umetsu,
  ``Higher-spin currents and thermal flux from Hawking radiation,''
  Phys.\ Rev.\  D {\bf 75} (2007) 124004
  [arXiv:hep-th/0701272];
  ``Fluxes of Higher-spin Currents and Hawking Radiations from Charged Black
  Holes,''
  Phys.\ Rev.\  D {\bf 76} (2007) 064015
  [arXiv:0705.3494 [hep-th]];
  ``Higher-spin Gauge and Trace Anomalies in Two-dimensional Backgrounds,''
  arXiv:0710.0453 [hep-th];
  ``Hawking Radiation via Higher-spin Gauge Anomalies,''
  Phys.\ Rev.\  D {\bf 77} (2008) 045007
  [arXiv:0710.0456 [hep-th]].
\bibitem{baner}
  R.~Banerjee and S.~Kulkarni,
  ``Hawking Radiation and Covariant Anomalies,''
  Phys.\ Rev.\  D {\bf 77} (2008) 024018
  [arXiv:0707.2449 [hep-th]];
  ``Hawking Radiation, Effective Actions and Covariant Boundary Conditions,''
  Phys.\ Lett.\  B {\bf 659} (2008) 827
  [arXiv:0709.3916 [hep-th]];
  S.~Kulkarni,
  ``Hawking Fluxes, Back reaction and Covariant Anomalies,''
  arXiv:0802.2456 [hep-th].
\bibitem{hayward} S.\ A.\ Hayward,
  ``Formation and evaporation of nonsingular black holes,''
  Phys.\ Rev.\ Lett.\ {\bf 96} (2006) 031103 [arXiv:gr-qc/0506126]
\bibitem{nss}
  P.\ Nicolini, A.\ Smailagic and E.\ Spallucci,
  ``Noncommutative geometry inspired Schwarzschild black hole,''
  Phys. Lett. B {\bf 632} (2006) 547
  [arXiv:gr-qc/0510112].
\bibitem{ss}
  A.\ Smailagic and E.\ Spallucci,
  ``Feynman path integral on the noncommutative plane,''
  J.\ Phys.\ A {\bf 36} (2003) L467
  [arXiv:hep-th/0307217];
  ``UV divergence-free QFT on noncommutative plane,''
  J.\ Phys.\ A {\bf 36} (2003) L517
  [arXiv:hep-th/0308193].
\bibitem{aw}
  L.~Alvarez-Gaume and E.~Witten,
  ``Gravitational Anomalies,''
  Nucl.\ Phys.\  B {\bf 234} (1984) 269.
\bibitem{bz}
  W.~A.~Bardeen and B.~Zumino,
  ``Consistent And Covariant Anomalies In Gauge And Gravitational 
  Theories,''
  Nucl.\ Phys.\  B {\bf 244} (1984) 421.
\bibitem{bk}
  R.~A.~Bertlmann and E.~Kohlprath,
  ``Two-dimensional gravitational anomalies, Schwinger terms and 
  dispersion relations,''
  Annals Phys.\  {\bf 288} (2001) 137
  [arXiv:hep-th/0011067].
\end{thebibliography}
\end{document}